# Melting of interference in the fractional quantum Hall effect: Appearance of neutral modes

R. Bhattacharyya, Mitali Banerjee[+], Moty Heiblum[*], Diana Mahalu, and Vladimir Umansky

Braun Center of Submicron Research, Department of Condensed Matter Physics, Weizmann Institute of Science, Rehovot 761001, Israel

([*] Corresponding author: moty.heiblum@weizmann.ac.il)

**Electrons living in a two-dimensional world under a strong magnetic field - the so-called fractional quantum Hall effect (FQHE) - often manifest themselves as fractionally charged quasiparticles (anyons). Moreover, being under special conditions they are expected to be immune to the environment, thus may serve as building blocks for future quantum computers. Interference of such anyons is the very first step towards understanding their anyonic statistics. However, the complex edge-modes structure of the fractional quantum Hall states, combined with upstream neutral modes, have been suspected to prevent an observation of the much sought after interference of anyons. Here, we report of finding a direct correlation between the appearance of neutral modes and the gradual disappearance of interference in a Mach-Zehnder interferometer (MZI), as the bulk filling factor is lowered towards Landau filling $\nu_B$ =1; followed by a complete interference quench at $\nu_B$=1. Specifically, the interference was found to start diminishing at $\nu_B$~1.5 with a growing upstream neutral mode, which was detected by a born upstream shot noise in the input quantum point contact (QPC) to the MZI. Moreover, at the same time a $\nu_{QPC}$ =1/3 conductance plateau, carrying shot-noise, appeared in the transmission of the QPC - persisting until bulk filling $\nu_B$ =1/2. We identified this conductance plateau to result from edge reconstruction, which leads to an upstream neutral mode. Here, we also show that even the particle-like quasiparticles are accompanied by upstream neutral modes, therefore suppressing interference in the FQHE regime.**

The discovery of the fractional quantum Hall effects (FQHE) had launched a four-decade long search for anyonic quasiparticles [1]. Unlike bosons and fermions, these quasiparticles are restricted to exist in only two dimensions. Their exchange statistics [2] is particularly of interest, as they are not expected to follow the ubiquitous Fermi-Dirac or Bose-Einstein statistics. States of the FQHE regime, and the edge modes that they support, are yet to be fully understood. What makes them elusive at times is the presence of different types of neutral modes [3], which are inert to standard electrical measurements. Indeed, neutral modes have found recently a niche in condensed matter physics; be topological abelian [4-7], or, the much sought after, the non-abelian type [4, 8, 9]. It worth mentioning also the non-topological neutral modes that emerge spontaneously due to unexpected edge-reconstruction [10].

Interference of anyons is one of the stepping-stones needed to establish their exchange statistics. In spite of substantial theoretical work [11-15], an observation of anyonic statistics still remains an experimental challenge. To achieve this purpose, electronic interferometers, be it a Mach-Zehnder interferometer (MZI) [16, 17] or a Fabry-Perot interferometer (FPI) [18, 19], have been realized in the QHE regime. While interference of electrons in the Integer QHE is relatively easy to find, reports on observation of interfering quasiparticles [20, 21] was not universally accepted. Our own efforts to observe quasiparticle interference in a MZI, as the bulk filling factor approached $\nu_B$=1 and lower, failed [22]. In this letter, we prove that proliferation of upstream neutral modes dephases quasiparticle interference. We found a direct correlation between the appearing of neutral modes in a MZI and the disappearing of the Aharonov-Bohm (AB) interference.

Three different GaAs-AlGaAs heterostructures, with 2D areal density $n$=(0.88-1.00)$\times10^{11}$cm$^{-2}$ and a 4.2K dark-mobility of $\mu$=(4.6-5)$\times10^{6}$cm$^{2}$/V-s, were used to fabricate our structures. Devices were patterned on a ‘wet-etched’ ‘Hall bar’ type mesas. The ohmic contacts are alloyed Ni/Ge/Au and the gates are made of an evaporated ~20nm thick PdAu/Au layers. The electrical conductance, as well as the shot noise, were measured using a two-stage amplification setup, composed of a cooled (to 4.2K) home-made voltage preamplifier followed by a commercial RT amplifier (NF-SA 220F5).

We adopted a device design to observe neutral modes and simultaneously observe the presence of AB interference. Figure 1 shows a schematic of the MZI (lithographic area ~30μm$^2$) with a preceding $QPC_0$ (see SEM image in Fig. S1a). Source contacts are labeled S1 and S2, drains D1-D3, cooled preamplifiers A, and grounds G (shorted to the 'cold finger' at ~10mK). Shot noise (in a partitioned $QPC_0$) is used in the determination of the electron temperature [23].

The MZI is composed of $QPC_L$ and $QPC_R$, serving as beam-splitters, and two outputs, D1 and D2 (here, grounded). The QPCs' gates are ~40nm wide (see SEM image in Fig. S1b). The modulation-gate (MG) can vary the threaded AB flux in The MZI. The source $S_N$, placed 7μm downstream from the preamplifier, is used to excite an upstream neutral mode (from its hot spot, see Fig. 1) [24].

When the bulk is set to a filling $\nu_B$, the profile of the chiral edge mode can be accessed by scanning the transmission of a QPC (from fully open to fully close). A fully open QPC, as well as the QPC tuned to a conductance plateau, indicate full transmission (and full reflection) of certain edge modes. Since no particles partitioning takes place, the transmitted current is noiseless [23]. However, we found an exceptional behavior in hole-conjugate states, $½<\nu_B<1$: the conductance plateaus in the transmission of a QPC carry downstream shot noise whose Fano-factor (see below) equals the bulk fractional filling. This type of *noise-carrying-plateau* had been identified before to result from an excitation of upstream neutral modes in the partly pinched QPC (referred to as *noisy-plateau* onwards) [25].

We dwell now on the correlation we found between AB interference and neutral modes. The *noisy-plateau* will be one of our 'canaries in a coal mine' for indicating the appearance of spontaneous edge-reconstruction, which gave birth to an excited upstream neutral mode. The *noisy-plateau* is characterized by a Fano factor, $F$, defined via the spectral density of the current fluctuations (at zero temperature): $S=2FeIt(1-t)$, where $I$ the impinging current, $t$ the transmission coefficient of the QPC, and $e$ the electron charge [26].

We start by monitoring the transmission through $QPC_L$ as the bulk filling is lowered from $\nu_B$=2 towards the fractional regime. Simultaneously, the AB interference of the outer-most $v$=1 edge mode in the MZI is monitored. Considerable AB oscillations were observed as the bulk filling approached $\nu_B$~5/3 (Fig. 2c – two top right panels). As obvious in Fig. 2, the oscillation visibility gradually diminished with lowering further the filling factor toward $\nu_B$=1; concomitantly with widened the $v$=1/3 *noisy-plateau* (see also Fig. S2). As shown in Figs. 3 & 4, the Fano factor was equal to the quantized bulk fillings ($\nu_B$=1,2/3,…..4/7) [26, 27].

A direct measurement of the upstream neutral mode emanating from a partly pinched QPC (transmission range was $t$=0 to $t$=1)) had been performed at $\nu_B$=1 by measuring the upstream noise in a contact placed 7μm upstream from the QPC along a gated mesa (Fig. S2a) [10]. The measured spectral density of the noise peaked at $S$~6x$10^{-30}A^2$/Hz in a wide range of transmissions (Fig. S2b). This measurement complements the measured shot-noise on the $v$=1/3 conductance plateau in a QPC at $\nu_B$=1.

We attribute the onset of the neutral mode as the bulk filling approaches $\nu_B$=1 to a spontaneous reconstruction of the edge potential that leads to two downstream chiral edge modes corresponding to fillings $v$=2/3 and $v$=1/3. The reconstruction of the edge at $\nu_B$ =2/3 was already demonstrated in Ref. 26, with two co-propagating $v$=1/3 modes were found to formed. A similar reconstruction takes place in $\nu_B$ =1 - resulting with two co-propagating modes, $v$=2/3 and $v$=1/3 (Fig. S3a). These two modes split at the QPC, to equilibrate later with the released energy excites an upstream neutral mode (Fig. S3b). Moving backwards (upstream), the neural modes fragment at the QPC to electron-hole pairs, giving rise to downstream shot noise (Fig. S4).

It is important to stress that the appearing of neutral modes at $\nu_B$ =1 is not topological, and they are relatively short lived. The thermal conductance of any bulk filling must obey a universal value; being for $\nu_B = 1$ a single quanta of thermal conductance [7]. As the bulk filling is lowered further, entering the ‘hole-conjugate regime’, the presence of counter-propagating modes (usually, hosting more than a single neutral mode) becomes a ‘topological must’ [7, 28, 29]. Indeed, the $v$=1/3 *noisy-plateau* remained prominent (accompanied by more *noisy-plateaus*) up to bulk filling $\nu_B$ =1/2. Figure 4 shows such plateaus and the corresponding shot noise on the $\nu$=1/3 plateau as the filling is being lowered ($\nu_B$=2/3 - $\nu$=1/3 plateau; $\nu_B$ =3/5 - $\nu$=1/3 & 2/5 plateaus; $\nu_B$ =4/7 - $\nu$=1/3 & 2/5 & 3/7 plateaus). The Fano factor (no shown) in every *noisy-plateau* was the same as the corresponding bulk filling ($F$=2/3, 3/5 and 4/7, respectively). As the bulk filling approached $\nu_B$=1/2, the $\nu$=1/3 *noisy-plateau* shrank (in span of the gate voltage) - disappearing completely at $\nu_B$=1/2. For bulk fillings lower than $\nu_B$ =1/2, *i.e.*, in the particle-like regime ($\nu_B = 4/9$, 3/7, 2/5), all the conductance plateaus in the QPC (corresponding to integer number of Composite Fermion modes) did not carry any downstream noise – as expected.

A natural question can be posed here regarding the particle-like fractional states, where topological edge-reconstruction and neutral modes are not expected. Yet, we found that upstream neutral edge modes are ubiquitous in these states, though they survive over shorter propagating distances (~10μm along a gated [10] or an etched mesa).

We tried to quench the non-topological upstream modes by a variety of 'gated-etched' edges. The source marked as $S_N$ (see Fig. 1) was used to excite (with DC bias) upstream neutral modes. The excitation energy was provided by the 'hot-spot' at the upstream side of $S_N$ [4, 24]. Excited upstream chiral neutral modes were found via an excess-noise measured in D1 (Fig. S5) [10]. The measured noise was much feeble than the noise measured under similar conditions at $\nu_B$=2/3, and it vanished at ~90mK. No upstream noise could be measured when the propagating distance was increased to 30μm at the lowest temperature. We failed to quench these modes by sharpening the edge-potential profile by positively biasing a side-gate on the mesa (Fig. S6).

With the onset of edge reconstruction at bulk filling $\nu_B$~1.5, the observed interference in a MZI decays, to fully quench at $\nu_B = 1$. The interference does not recover again in the fractional regime, $\nu_B < 1$. This is a direct result of the very existence of neutral modes, be it topological or resulting from edge-reconstruction. The exact cause of edge-reconstruction in the lowest Landau level, *i.e.*, $\nu_B < 2$, leading to proliferation of upstream neutral modes, is not clear, but is assumed to result from the softness of the edge potential. Artificial construction of chiral fractional modes away from the physical edge of the 2DEG may lead to a more controlled local potential and thus quench non-topological neutral modes.

## Author contributions

R.B., M.B., and M.H. designed the experiment. R.B. and M.B. fabricated the structures, performed the measurements, and did the analysis. V.U. grew the actual 2DEG heterostructures. All contributed to the write up of the manuscript.

## Acknowledgements

We thank Yuval Gefen, Yunchul Chung and Hiroyuki Inoue for insightful discussions. M.H. acknowledges the partial support of the Israeli Science Foundation (ISF), the Minerva foundation, and the European Research Council under the European Community's Seventh Framework Program (FP7/2007– 2013)/ERC Grant agreement 339070.

[+]Current address: Department of Physics, Columbia University, New York, NY, USA.

**Figure Captions**

**Figure. 1| Device schematic.** Schematic of the structure. The outer periphery is defined by an etched mesa. The MZI is defined by two QPCs, labeled as $QPC_L$ and $QPC_R$, with the upper path separated by an etched mesa (a white arc). The modulation-gate, affecting the area of the MZI is labeled as MG. The path of the interfering edge-channel along the mesa is shown in black broken lines with arrows defining the chirality. Shot-noise measurements were performed using $QPC_0$ whereas for conductance measurements all the QPCs were used. $QPC_0$ can also be used to switch between $I_{S1}$ and $I_{S2}$ as the impinging edge channels on the MZI.

**Figure. 2| Correlation between the appearance of a $\boldsymbol{\nu = 1/3}$ *noisy-plateau* and the diminishing interference in the MZI.** a) A 'two-probe' quantum Hall resistance as a function of magnetic field in bulk filling $\nu_{\mathrm{B}} = 2$ to $\nu_{\mathrm{B}} = 1$. Filling factors corresponding to observed plateaus are noted with arrows showing the plateaus. The colored squares show the places where the experiments for b) and c) were performed. b) Plots show the conductance vs gate voltage applied to $QPC_L$ at different bulk states. The dotted magenta lines are guide to an eye of the $\nu = 1/3$ conductance plateau in the QPC. c) Plots show the corresponding interference signal of the outermost edge mode as a function of MG voltage in the MZI. Note the diminishing visibility of interference once 1/3 plateau in the QPC appears. The interference quenches once the plateau is fully-grown at $\nu_{\mathrm{B}} = 1$.

**Figure 3| Differential conductance and shot noise at $\boldsymbol{\nu_{\mathrm{QPC}} = 1/3}$ at bulk filling $\boldsymbol{\nu_B = 1}$.** The peripheral axes are defined for transmission through $QPC_0$ as a function of gate voltage. A plateau at $t$=1/3 (dotted magenta line) shows the formation of $e^2/3h$ conductance plateau in the QPC. a) The non-linear differential conductance of $QPC_0$ as a function of DC-bias applied to the source at $t$=1/3. b) Shot noise at the $t$=1/3 plateau. Black solid line shows the fitting curve to the shot-noise data. Partitioned quasiparticles charge (here, the Fano factor) and temperature as obtained by the fitting curve is provided as inset text.

**Figure. 4| Differential conductance and shot-noise in the hole-conjugate states $\boldsymbol{\nu_B = \frac{2}{3}, \frac{3}{5}, \frac{4}{7}}$.** Transmission through $QPC_0$ as a function of voltage on the gates showing conductance plateaus for three different filling factors. The insets show the excess noise measured on the v=1/3 plateau in each filling factor. Fitted lines through the noise data gives the Fano factors corresponding to each bulk filling.

**Melting of interference in the fractional quantum Hall effect: Appearance of neutral modes**

R. Bhattacharyya, Mitali Banerjee, Moty Heiblum*, Diana Mahalu, and Vladimir Umansky

Braun Center of Submicron Research, Department of Condensed Matter Physics, Weizmann Institute of Science, Rehovot 7610001, Israel

(* Corresponding author: moty.heiblum@weizmann.ac.il)

## Supplementary Information

### S1. SEM image of MZI device and 40 nm wide QPC used in all our experiments

Figure S1(a) is an SEM image of the device we have used in our experiments (See main text Fig. 1 for the schematic). Figure S1(b) shows an SEM image of a typical QPC, ~40nm wide, that has been used in all our experiments unless otherwise specified. The 1/3 conductance plateau in the QPC, although is well-visible in regular QPCs in hole-conjugate states, only such thin ~40nm wide QPCs gave rise to reproducible observation of 1/3 conductance plateau at $\nu_B = 1$. We suspect that only such thin QPCs provide enough resolution in terms of QPC-voltage to observe the small 1/3 conductance plateau at $\nu_B = 1$.

### S2. 2D map of interference visibility strongly correlated with 1/3 plateau in QPC

Figure S2 is an extension of Fig. 2. in the main text, that illustrates the interference visibility as a function of both magnetic field (B) and modulation-gate voltage ($V_{MG}$) . The 2D plot clearly shows fading of the visibility all the way to zero as $\nu_B = 1$ was approached. This fading hints at a possible first order phase transition taking place at the QPCs of the MZI due to the emergence of upstream neutral mode causing decoherence of the wave-function of partitioned electron, and thereby a quench of interference.

### S3. Direct measurement of upstream neutral mode at $\nu_B = 1$ generated in a QPC

We adopted another device with a regular QPC (~200 nm wide) to perform a direct measurement of upstream neutral mode generated in a QPC. Fig. S3a shows the device schematic, where an ohmic contact labelled 'D2' is placed very close to the QPC in the upstream direction to detect the neutral via noise measurement. A neutral mode ignited in the QPC upon biasing S, if allowed to propagate in the upstream direction due to edge reconstruction along the soft gate-defined potential, should be incident on ‘D2’ and cause a voltage-fluctuation resulting a voltage-noise. Fig. S3b presents the upstream noise measured in D2 at $\nu_B = 1$ as a function of QPC-transmission calibrated via differential transmission measurement using D1. Such an upstream noise could not be measured at $\nu_B = 2$. The direct measurement of upstream noise excited in a pinched QPC at $\nu_B = 1$ further strengthens our argument about the interference quenching due to emergent upstream neutral mode, manifested by 1/3 plateau in a QPC and shot-noise measured on the plateau. Note that in a similar device, upstream noise was measured in all fractional states including the particle-like states [10].

## S4. A possible explanation of 1/3 conductance plateau at $\nu_B = 1$ : edge-reconstruction model

The 1/3 conductance plateau at $\nu_B = 2/3$ [26] has been explained in the Wang-Meir-Gefen [30] model in terms of edge reconstruction. The left figure of Fig. S4(a) shows the edge structure model proposed for $\nu_B = 2/3$. Equilibration between two upstream 1/3 mode and a downstream 1 mode results into a downstream 1/3 channel and two upstream neutrals. The farthest downstream 1/3 mode remaining un-equilibrated, continues as a downstream 1/3 mode, thereby resulting into two downstream 1/3 mode (as observed in the transmission through QPC experiments) and two upstream neutral modes. This model also satisfies the requirement of thermal conductivity $\kappa = 0$ for $\nu_B = 2/3$, which is a measure of net number of edge modes in a quantum Hall system and has been verified experimentally [7].

We extend this model to $\nu_B = 1$ to explain the 1/3 conductance plateau and upstream neutral mode observed in our experiments. The right figure of Fig S4(b) shows our proposed model for $\nu_B = 1$ including a possible edge-reconstruction very similar to $\nu_B = 2/3$. In this case, equilibration between downstream 1 and upstream 1/3 mode can give rise to a net downstream 2/3 mode and upstream neutral. The farthest downstream 1/3 mode being un-equilibrated, continues and results into a downstream 2/3, a downstream 1/3 and an upstream neutral mode. Since this upstream neutral mode is not originated topologically, one would expect its strength to be much fainter than $\nu_B = 2/3$ where one of the two neutral modes must originate topologically. This indeed is in accordance with our observation that has been shown in Fig. S3. This model also satisfies κ = 1 for $\nu_B = 1$ that has also been verified experimentally [7].

Figure S4(b) shows a simple model to understand the origin of observed shot-noise on the 1/3 conductance plateau intuitively. Once the QPC is set on the 1/3 conductance plateau, an equilibration between hot edge and cold edge modes at different chemical potential can be anticipated, giving rise to a neutral mode moving in the upstream direction, towards the QPC. The neutral mode, upon arrival in the QPC and partitioned there, can give rise to downstream shot-noise [26]. This simple model does not quantify the observable shot-noise and does not explain the quantization of the Fano-factor (see main text), but we hope an extension of the 'neutralon' model in ref. [25] can explain them.

## S5. Upstream neutral mode at $\nu_B = 1/3$ along etched mesa over short distance

As has been mentioned in the main text, evidence of upstream neutral mode in particle-like state, *e. g.*, $\nu_B = 1/3$ has been found while sourcing $S_N$ and measuring noise upstream at D1. A comparison has also been shown in Fig. S2 between the upstream neutral (a) at $\nu_B = 1/3$ and the same (b) at $\nu_B = 2/3$, maintaining same distance between $S_N$ and D1. While the latter is much stronger in magnitude being topologically protected, the weak neutral at $\nu_B = 1/3$ propagating even along etched-mesa defined sharp-edge potential is evidence of universal appearance of neutral modes in fractional quantum hall regime in shorter distances. Figure S5(c) illustrates the temperature dependence of the neutral mode $\nu_B = 1/3$. It therefore strengthens our argument of dephasing caused by neutral modes, based on observed correlation between the two.

## S6. Side-gate on a mesa to minimize edge-reconstruction in fractional quantum Hall regime by sharpening edge-potential

Here we propose an experiment aimed at sharpening the edge-potential even more than what can usually be obtained at the edge of an etch-defined mesa. The motivation of the proposal is the observation of upstream neutral along etch defined mesa over short distance at $\nu_B = 1/3$, possibly due to edge-reconstruction due to soft edge-potential. A sharper edge-potential in such case would lead to minimum, at best, zero edge-reconstruction resulting in abolishment of such short-lived non-topological neutral modes.

Fig. S6 (a) shows the schematic of the experiment, where a side-gate along the vertical wall of the etched-mesa extending up to the depth of the 2-DEG is placed. Intuitively one can deduce that an application of positive gate-voltage on the side-gate should stretch the 2-DEG closer to the mesa-boundary, and therefore, should lead to a sharper fall of the 2-DEG density-profile. Upon application of a proper positive bias on the side-gate, the sharpest possible 2-DEG density-profile can therefore be obtained, where edge-reconstruction should be minimum or even non-existent. We have numerically solved self-consistent Schrodinger-Poisson equation in two dimension for such a system, using a 2D Schrodinger-Poisson solver AQUILA (free Matlab toolbox) using appropriate boundary conditions (Dirichlet or Neumann). Figure S6 (b) shows the result of the simulation for such a system. With zero bias on the side-gate, we obtain the density profile provided only by an etch-defined mesa. With increasing positive bias on the side-gate, the 2-DEG density profile closer to the side-gate approaches the mesa-boundary that results into sharpening of the density profile at the edge. At a certain bias voltage dependent on the details of the growth-parameters, we obtain the sharpest possible density-profile at the edge, which should contain minimum edge-reconstruction in the fractional quantum Hall regime. Although our experimental trials (twice) did not come up with useful results, we think upon circumvention of the fabrication-related challenges to develop a reproducible side-gated mesa might lead to partial or complete eradication of unwanted edge-reconstruction in the fractional quantum Hall regime.

**Figure Captions**

**Fig. S1| SEM image of device used for the experiments..** a) False colored SEM image of the device consisting of an MZI and $QPC_{0.}$ The gates along with air-bridges are shown in green and the grounded ohmic contact inside MZI is shown in yellow. b) A ~40nm wide QPC defined by 20nm thick PdAu/Au evaporation. While using double-layer resist for e-beam lithography, a cold temperature development (~$13^0$ C) is recommended for obtaining such sharp features in the design.

**Fig. S2| MZI interference visibility as function of magnetic field and modulation-gate voltage between $\nu_B = 2$ and $\nu_B = 1$.** The 2D plot shows a fast Fourier transform of the best interference signal as a function of modulation-gate voltage ($V_{MG}$) periodicity obtained from the MZI at different magnetic fields from $\nu_B = 2$ to $\nu_B = 1$. The color encodes the visibility of the interference signal with a color-bar shown in the right. As a reference to the reader, a 2-probe Hall resistance data from $\nu_B = 2$ to $\nu_B = 1$ is shown in green with axis labelled in the right. Two vertical dotted lines divides the whole 2D map into three segments, each containing a typical conductance measurement with $QPC_L$ as inset. The dotted magenta lines in the inset shows 1/3 $\frac{e^2}{h}$ conductance through the QPC. As pointed out in Fig. 2 in the main text, the visibility drops once the 1/3 conductance plateau appears. Interference quenches once the 1/3 conductance plateau gains its full strength.

**Fig. S3| Direct measurement of upstream noise generated in a QPC at $\nu_B = 1$.** a) Device schematic where ohmic contacts are shown in yellow and the QPC is shown in green. Ohmic contact labelled 'D2' is placed just beside one of the gates defining the point-contact, 7μm away in the upstream direction from the point contact. Ohmic contact 'S' is biased to measure voltage-noise in D2. Chiral charge modes are drawn in 'black', upstream neutral generated in the QPC is drawn in 'red'. b) Current noise due to upstream neutral detected in D2 as a function of the transmission of QPC from fully open to fully closed, measured with a bias current of 5nA applied to S. The inset shows the upstream noise as a function of bias current applied to S for two different QPC transmissions, $t$ = 0.94 and $t$ = 0.53.

**Fig S4| Possible explanation of upstream neutral mode and shot-noise on 1/3 conductance plateau.** a) Left figure shows proposed edge structure at $\nu_B = 2/3$ for soft edge-potential [30]. Right figure shows our proposed edge structure at $\nu_B = 1$ for soft edge-potential as an extension of Wang-Meir-Gefen model. The downstream 1/3 mode closest to the physical edge does not equilibrate being far away from the upstream 1/3 mode, and therefore shows up as a 1/3 conductance plateau through QPC. Equilibration between downstream 1 and upstream 1/3 mode results into downstream 2/3 charge mode and an upstream neutral mode, which we detected in our setup (Fig. S3). b) An intuitive model to understand shot-noise on the 1/3 conductance plateau in the QPC. Source ohmic 'S' sources an edge channel at an elevated chemical potential,

that breaks into two co-propagating 1/3 and 2/3 modes when the QPC-voltage is set to be on the 1/3 conductance plateau. Both the transmitted 1/3 mode and reflected 2/3 mode mix with two other edges emanating from grounded ohmic contacts, therefore at zero chemical potential. Equilibration between the two edges after the QPC leads to upstream neutral mode that reaches the QPC. Being partitioned in the QPC to charge modes, the neutral mode generates downstream shot-noise.

**Fig. S5| Upstream neutral mode measured at $\boldsymbol{\nu_B = 1/3}$.** a) Upstream noise measured at D1 as a function of DC-bias current sourced in $S_N$ , 7 μm away from D1 along the mesa at $\nu_B = 1/3$. b) Same measurement performed at $\nu_B = 2/3$. Note that the noise is about 2 order of magnitude stronger than a). c) The upstream noise is measured at different temperature, from 20mK to 120mK, and has been plotted with vertical offset to show the gradual change in the noise profile as temperature is varied. At ~90mK the upstream noise measured becomes negligible. d) A vertical cut taken from c) at a DC-bias of 4nA. Temperature dependence of the upstream noise amplitude is quite linear.

**Fig. S6| Proposed schematic of side-gated mesa along with simulation results.** a) Schematic of a side-view of an etched mesa defined by black solid line with a side-gate shown in green. As a typical example, the 2DEG here is buried 120nm below the surface with a δ – doping layer spaced by 80nm. b) 2D Schrodinger-Poisson simulation of the structure shown in **a)** results into the following electron-density profile for different values on the side-gate denoted by $V_{side\text{-}gate}$. The origin of the plot is taken at the left mesa boundary of a). A depletion of the 2DEG for the first ~400nm from the mesa-boundary causes due to surface states in the mesa boundary [31]. As $V_{side\text{-}gate}$ is increased, a sharpening of the electron density profile in the right side of the mesa is observed. $V_{side\text{-}gate}$ cannot be increased too much since it will cause occupancy of surface states in the right side of the mesa-wall, causing again a soft edge-potential that one would like to avoid.

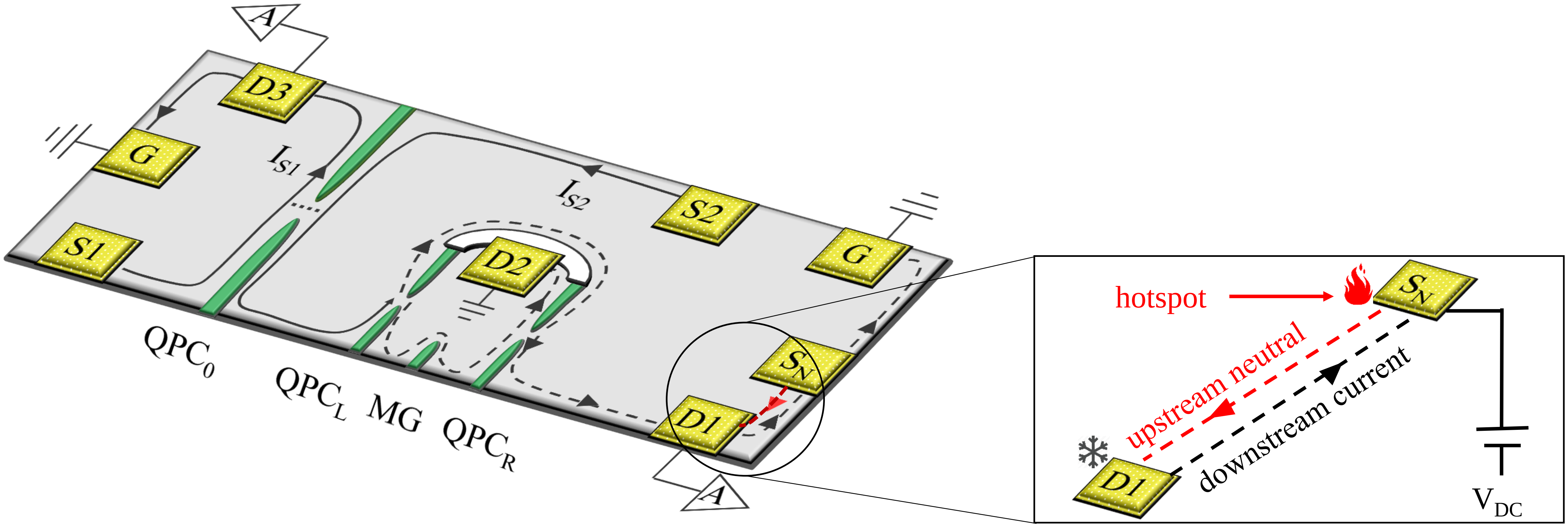

Fig 1

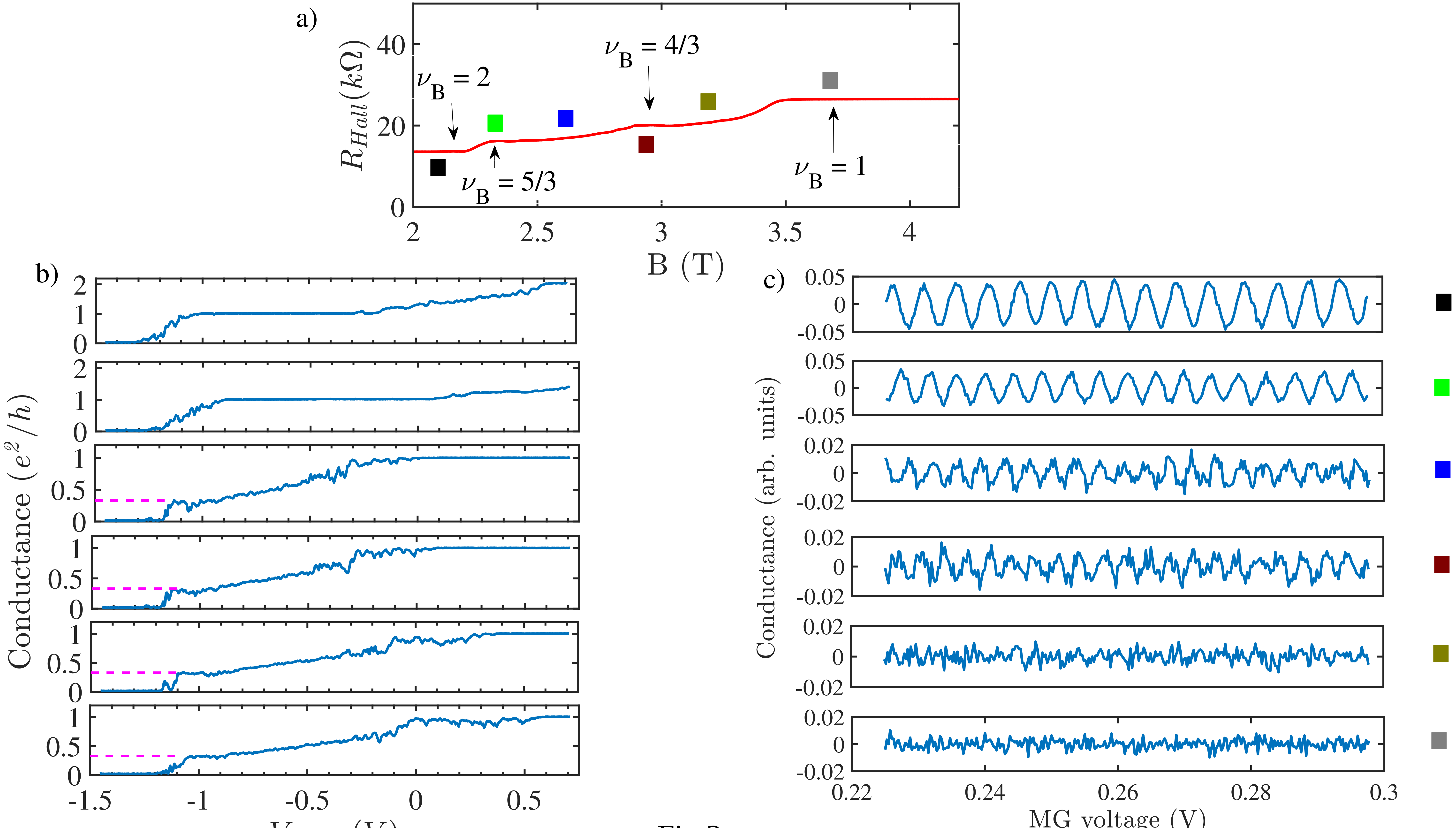

Fig 2

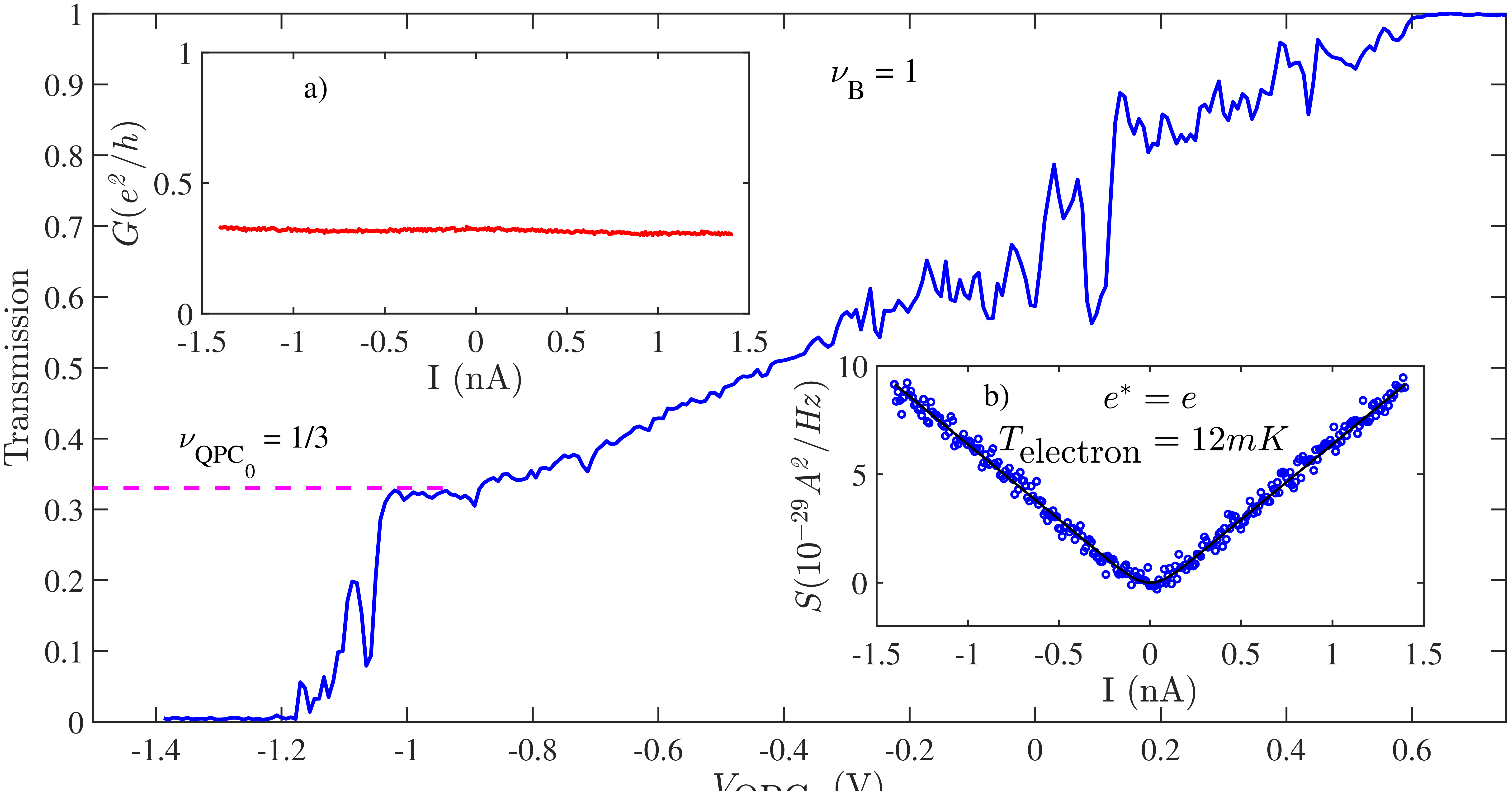

Fig 3

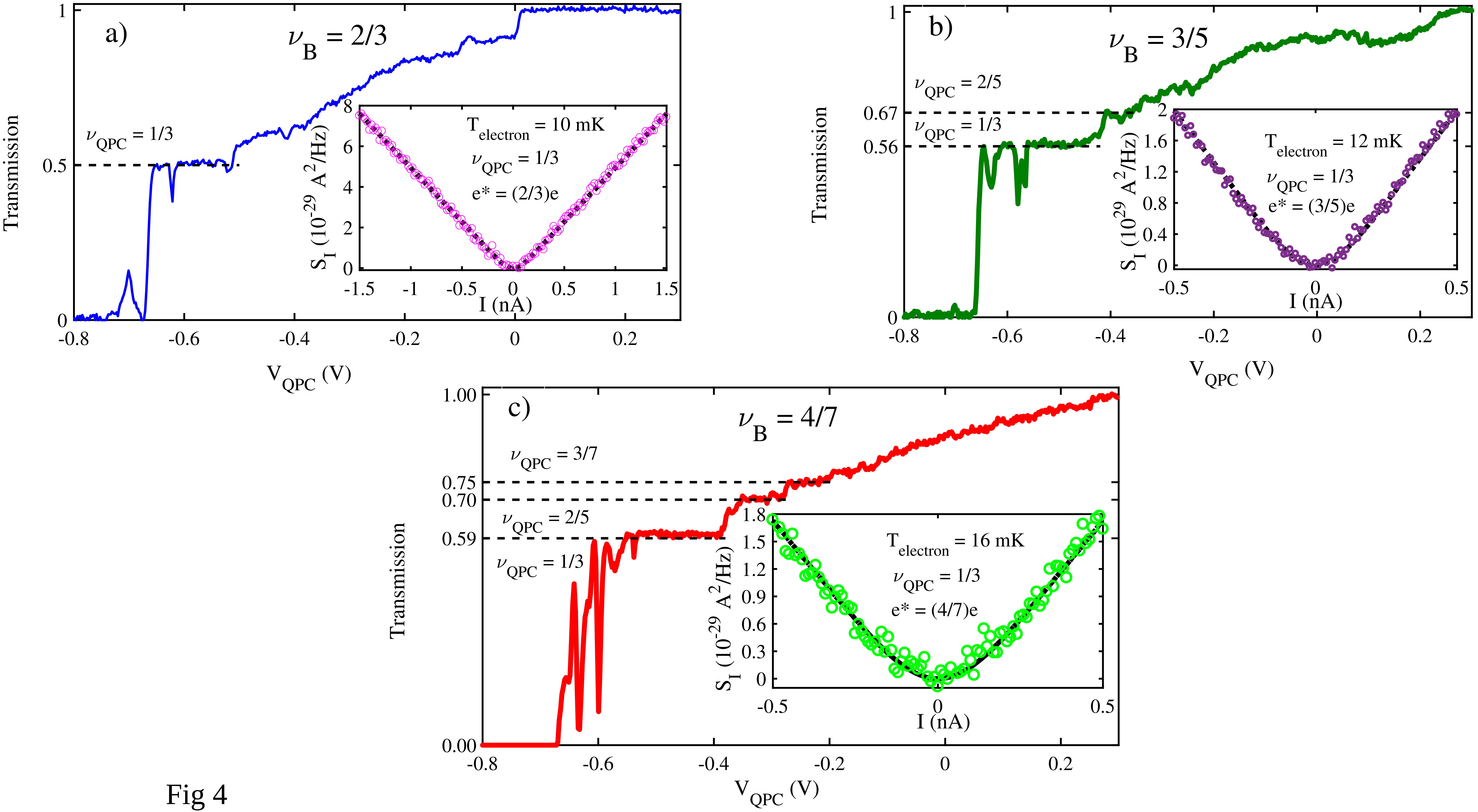

Fig 4

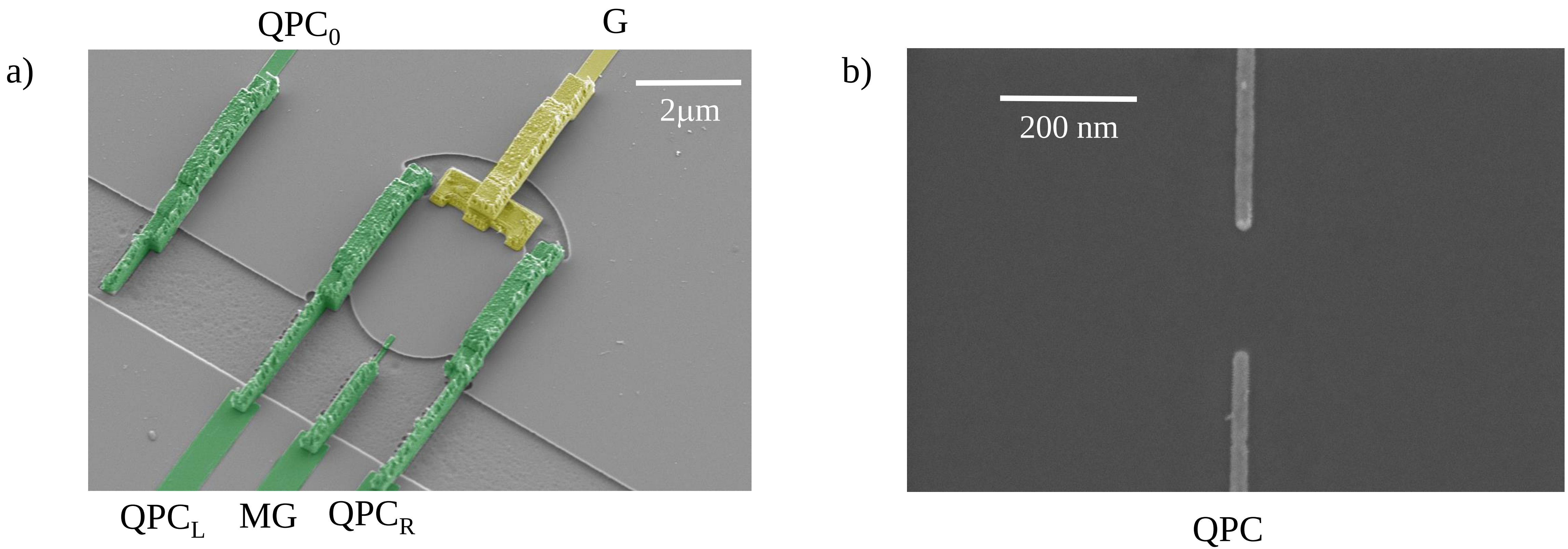

Fig S1

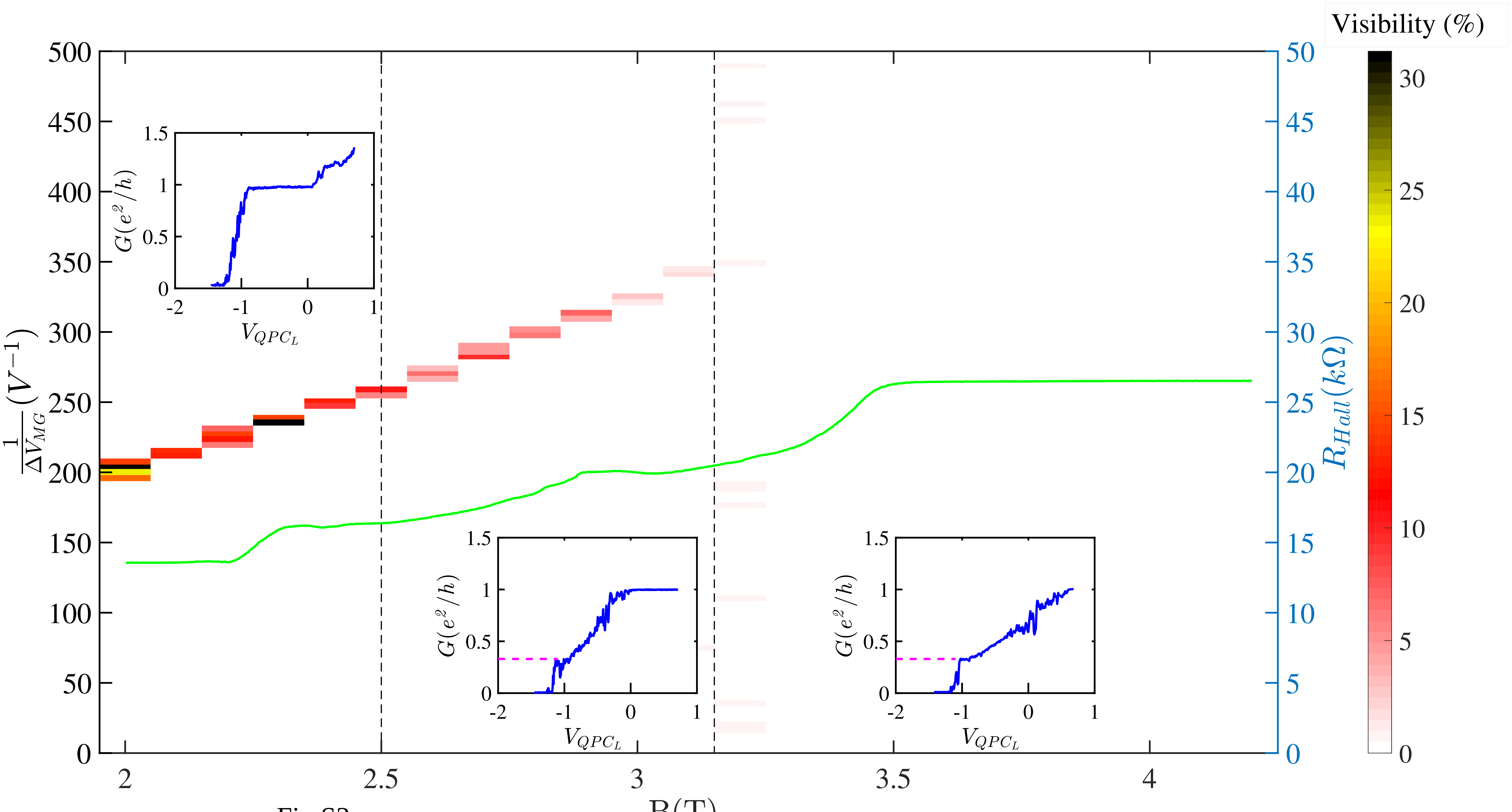

Fig S2

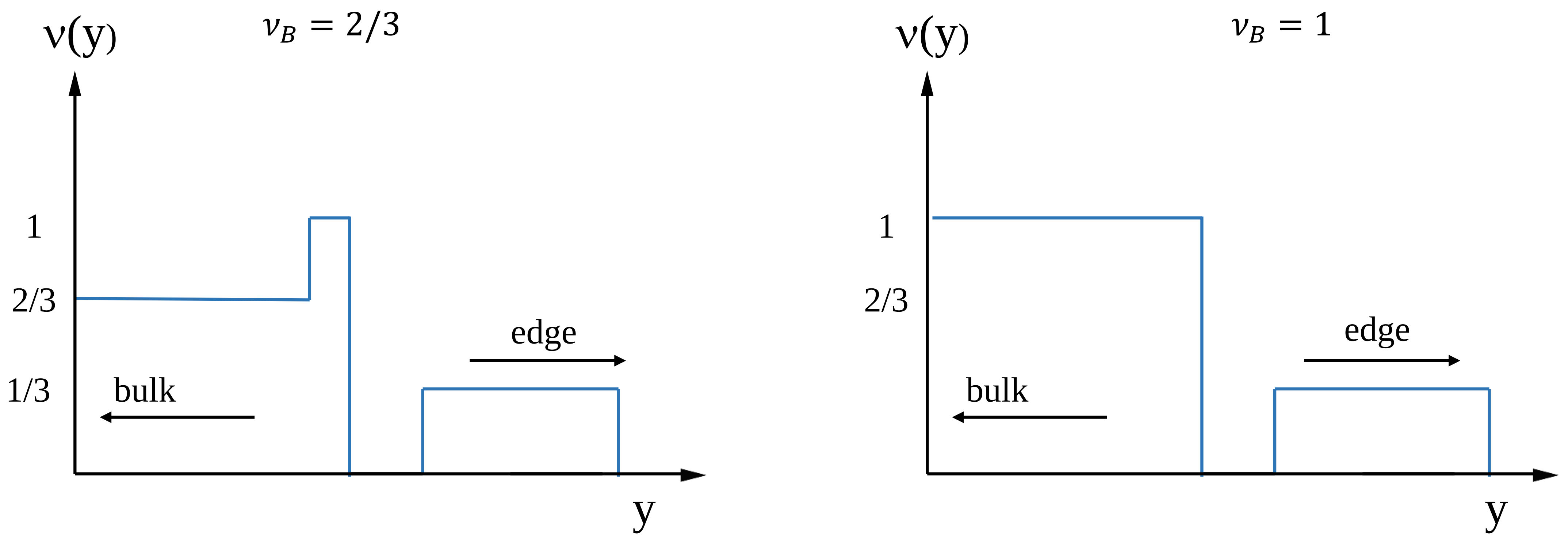

Fig S3a

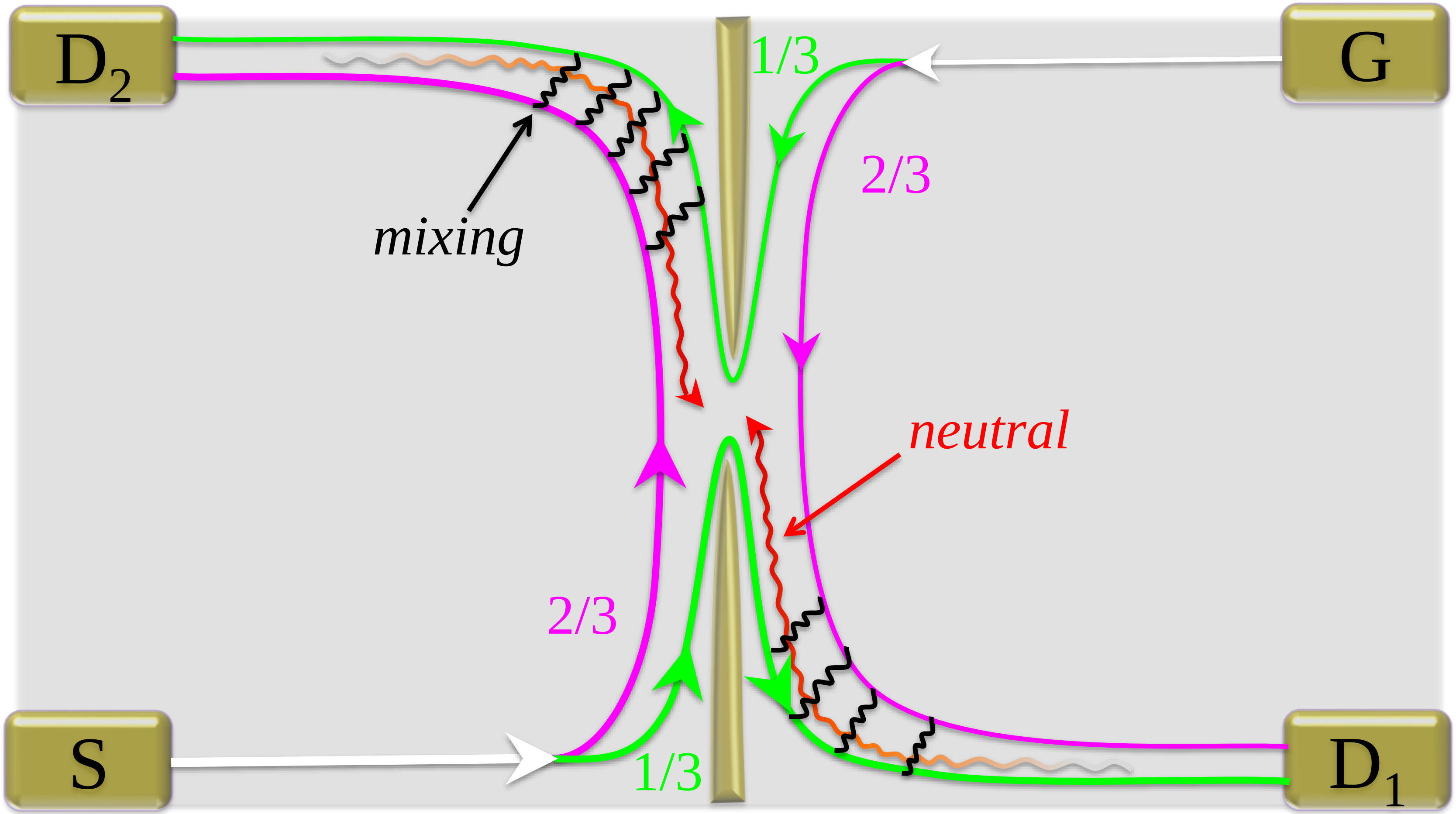

Fig S3b

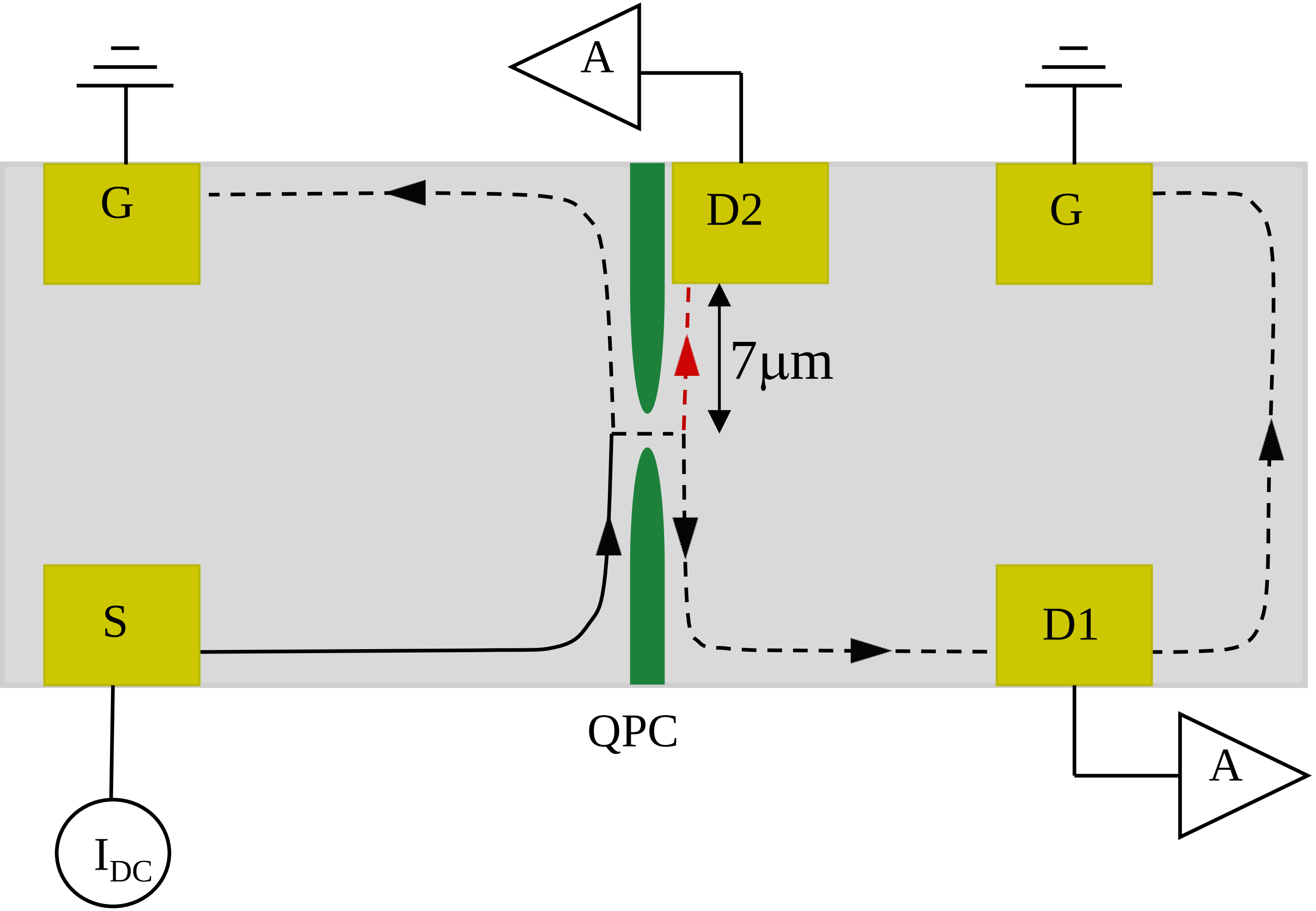

Fig S4a

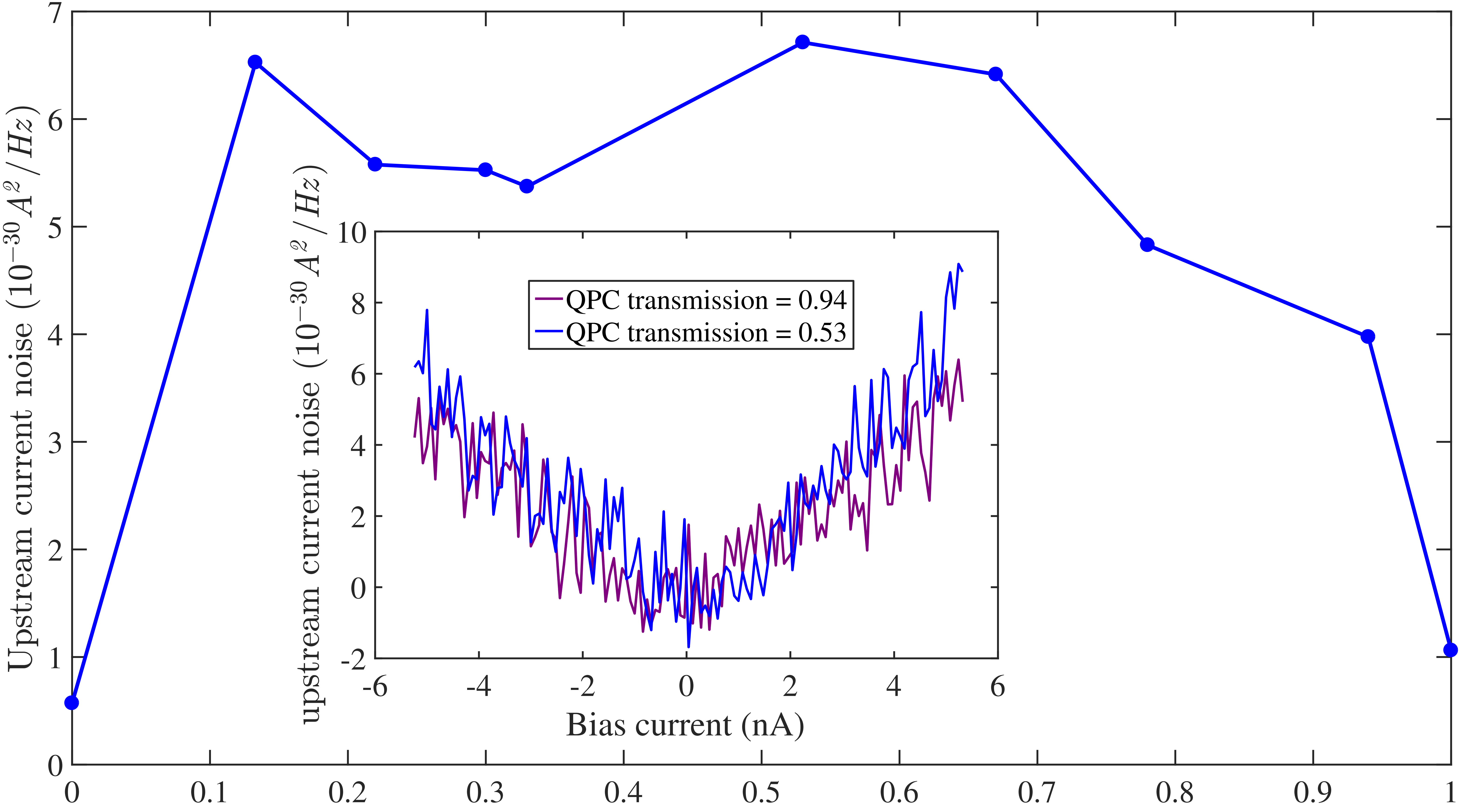

Fig S4b

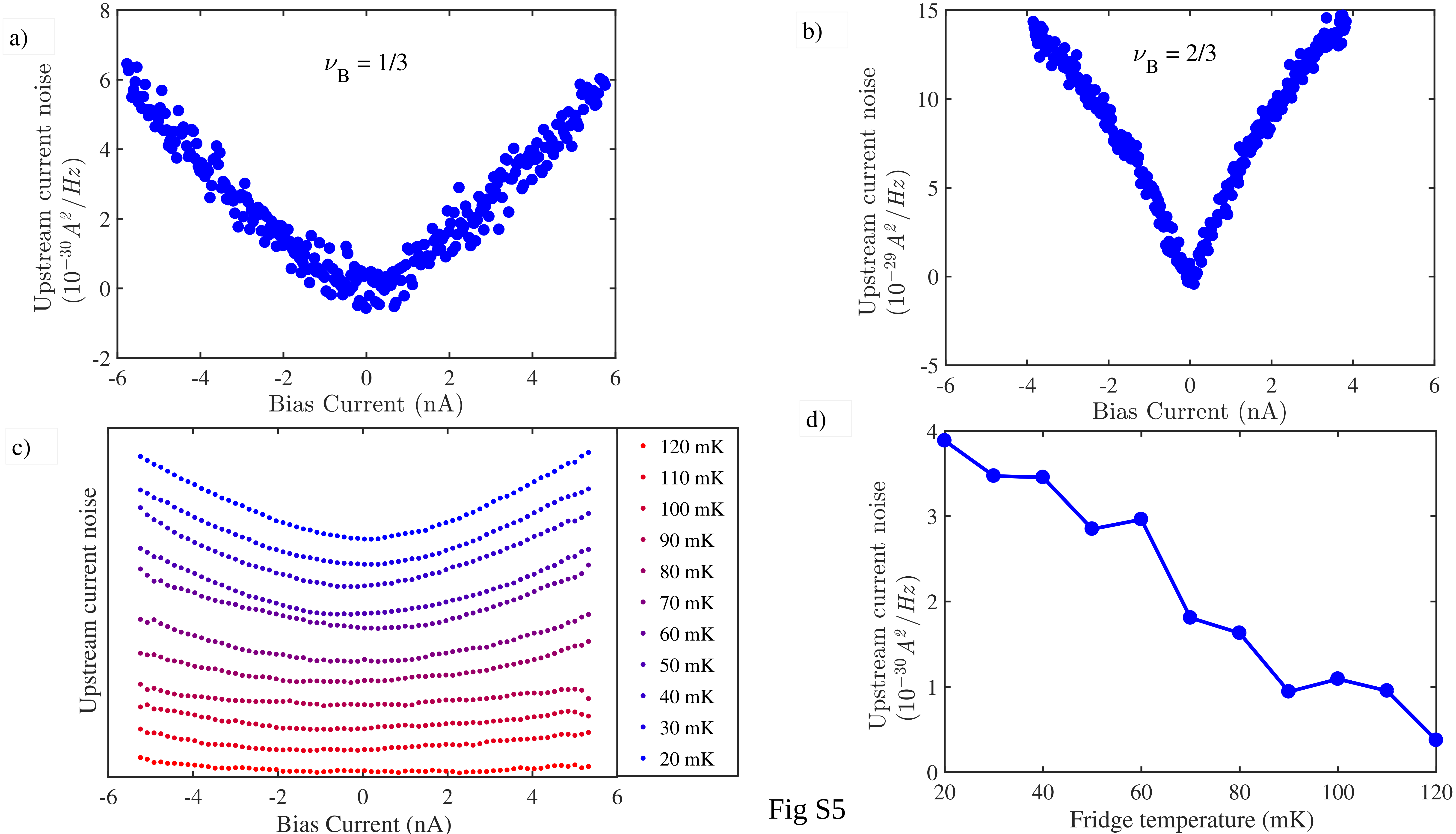

Fig S5

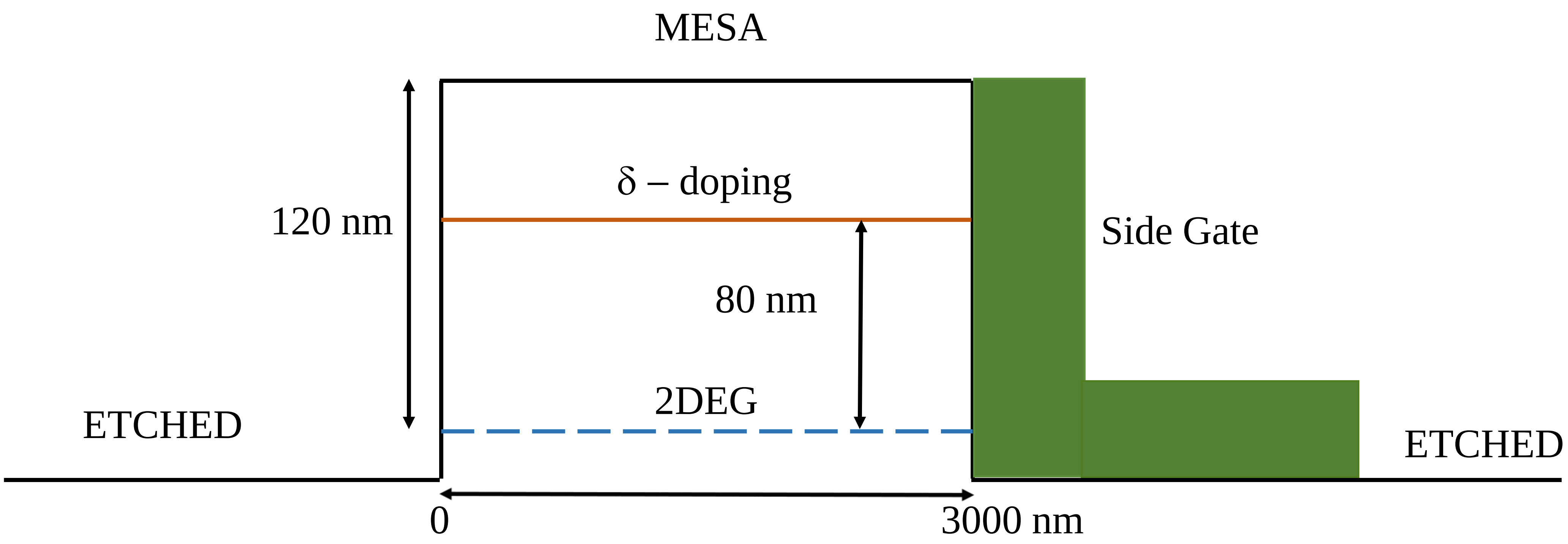

Fig S6a

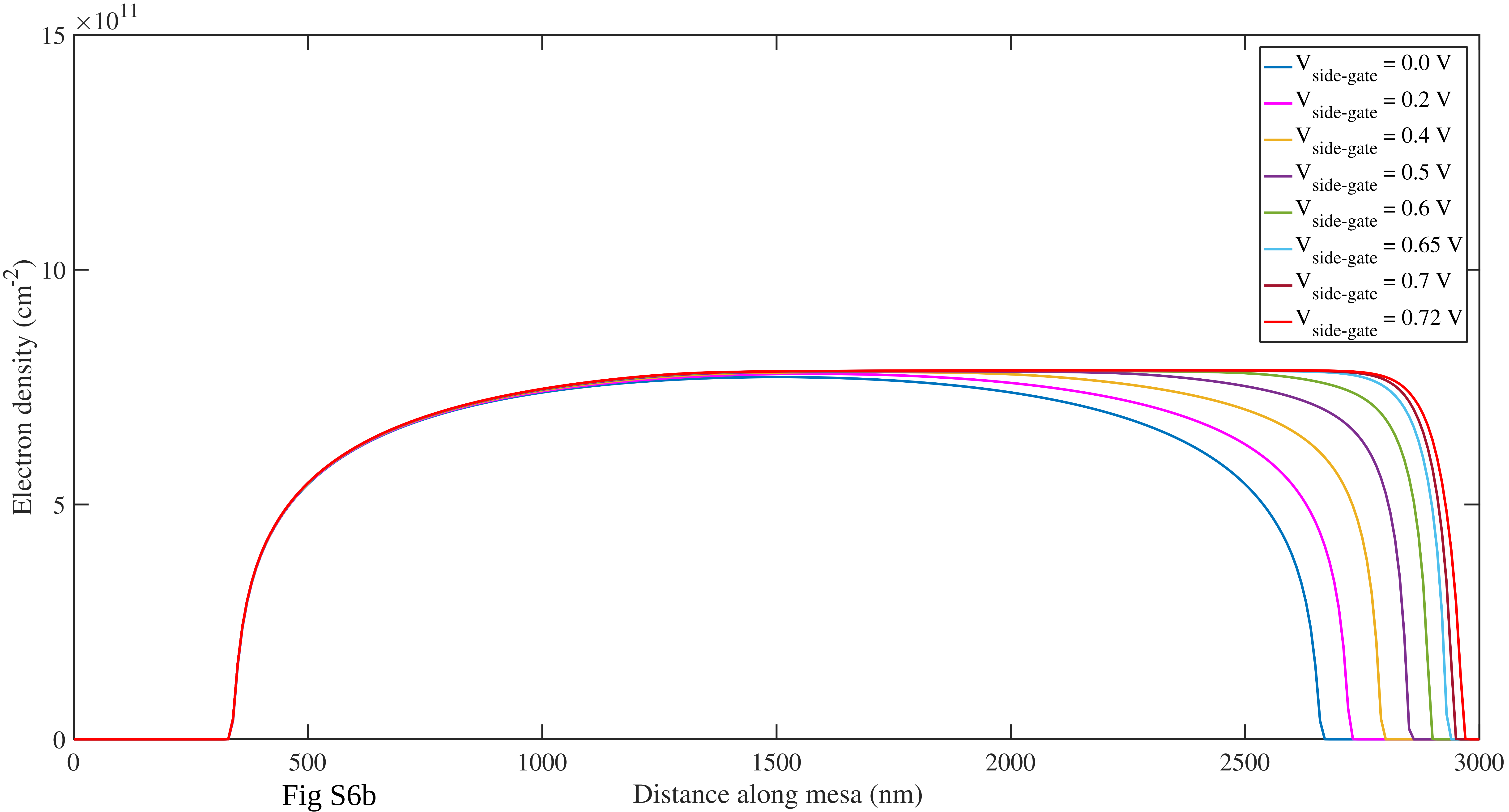
×10^11
15
10
5
0
Electron density (cm^-2)
0
500
1000
1500
2000
2500
3000
Distance along mesa (nm)
V side-gate = 0.0 V
V side-gate = 0.2 V
V side-gate = 0.4 V
V side-gate = 0.5 V
V side-gate = 0.6 V
V side-gate = 0.65 V
V side-gate = 0.7 V
V side-gate = 0.72 V
Fig S6b